\documentclass[11pt]{article}
\usepackage{graphicx}
\usepackage{amssymb,amsmath,amsfonts}

\def\Re{{\cal R \mskip-4mu \lower.1ex \hbox{\it e}\,}}
\def\Im{{\cal I \mskip-5mu \lower.1ex \hbox{\it m}\,}}
\def\ie{{\it i.e.}}
\def\eg{{\it e.g.}}
\def\etc{{\it etc}}
\def\etal{{\it et al.}}

\def\sub#1{_{\lower.25ex\hbox{$\scriptstyle#1$}}}
\def\tev{\,{\ifmmode\mathrm {TeV}\else TeV\fi}}
\def\gev{\,{\ifmmode\mathrm {GeV}\else GeV\fi}}
\def\mev{\,{\ifmmode\mathrm {MeV}\else MeV\fi}}
\def\mpl{\ifmmode M_{pl}\else $M_{pl}$\fi}
\def\mpl{\ifmmode \overline M_{Pl}\else $\bar M_{Pl}$\fi}
\def\to{\rightarrow}

\def\subw{_{\rm w}}
\def\mh{\ifmmode m\sbl H \else $m\sbl H$\fi}
\def\mch{\ifmmode m_{H^\pm} \else $m_{H^\pm}$\fi}
\def\mt{\ifmmode m_t\else $m_t$\fi}
\def\mc{\ifmmode m_c\else $m_c$\fi}
\def\mz{\ifmmode M_Z\else $M_Z$\fi}
\def\mw{\ifmmode M_W\else $M_W$\fi}
\def\mws{\ifmmode M_W^2 \else $M_W^2$\fi}
\def\mhs{\ifmmode m_H^2 \else $m_H^2$\fi}   
\def\mzs{\ifmmode M_Z^2 \else $M_Z^2$\fi}
\def\mts{\ifmmode m_t^2 \else $m_t^2$\fi}
\def\mcs{\ifmmode m_c^2 \else $m_c^2$\fi}
\def\mchs{\ifmmode m_{H^\pm}^2 \else $m_{H^\pm}^2$\fi}
\def\ztwo{\ifmmode Z_2\else $Z_2$\fi}
\def\zone{\ifmmode Z_1\else $Z_1$\fi}
\def\mtwo{\ifmmode M_2\else $M_2$\fi}
\def\mone{\ifmmode M_1\else $M_1$\fi}
\def\tb{\ifmmode \tan\beta \else $\tan\beta$\fi}
\def\xw{\ifmmode x\subw\else $x\subw$\fi}
\def\ch{\ifmmode H^\pm \else $H^\pm$\fi}
\def\lum{\ifmmode {\cal L}\else ${\cal L}$\fi}
\def\inpb{\,{\ifmmode {\mathrm {pb}}^{-1}\else ${\mathrm {pb}}^{-1}$\fi}}
\def\infb{\,{\ifmmode {\mathrm {fb}}^{-1}\else ${\mathrm {fb}}^{-1}$\fi}}
\def\epem{\ifmmode e^+e^-\else $e^+e^-$\fi}
\def\ppb{\ifmmode \bar pp\else $\bar pp$\fi}
\def\bsg{\ifmmode B\to X_s\gamma\else $B\to X_s\gamma$\fi}
\def\bsll{\ifmmode B\to X_s\ell^+\ell^-\else $B\to X_s\ell^+\ell^-$\fi}
\def\bstt{\ifmmode B\to X_s\tau^+\tau^-\else $B\to X_s\tau^+\tau^-$\fi}
\def\lamt{\ifmmode \tilde\lambda\else $\tilde\lambda$\fi}
\def\shat{\ifmmode \hat s\else $\hat s$\fi}
\def\that{\ifmmode \hat t\else $\hat t$\fi}
\def\uhat{\ifmmode \hat u\else $\hat u$\fi}

\newskip\zatskip \zatskip=0pt plus0pt minus0pt
\def\matth{\mathsurround=0pt}
\def\lsim{\mathrel{\mathpalette\atversim<}}

\def\atversim#1#2{\lower0.7ex\vbox{\baselineskip\zatskip\lineskip\zatskip
  \lineskiplimit 0pt\ialign{$\matth#1\hfil##\hfil$\crcr#2\crcr\sim\crcr}}}

\def\grtsim{\,\,\rlap{\raise 3pt\hbox{$>$}}{\lower 3pt\hbox{$\sim$}}\,\,}
\def\lsim{\,\,\rlap{\raise 3pt\hbox{$<$}}{\lower 3pt\hbox{$\sim$}}\,\,}

\renewcommand{\thefootnote}{\fnsymbol{footnote}}

\begin{document} \begin{titlepage}
\rightline{\vbox{\halign{&#\hfil\cr
&SLAC-PUB-11666\cr
}}}
\begin{center}
\thispagestyle{empty} \flushbottom { {
\Large\bf Higher Curvature Gravity in TeV-Scale Extra Dimensions
\footnote{Work supported in part
by the Department of Energy, Contract DE-AC02-76SF00515}
\footnote{e-mail:
$^a$rizzo@slac.stanford.edu}}}
\medskip
\end{center}

\centerline{Thomas G. Rizzo$^{a}$}
\vspace{8pt} 
\centerline{\it Stanford Linear
Accelerator Center, 2575 Sand Hill Rd., Menlo Park, CA, 94025}

\vspace*{0.3cm}

\begin{abstract}
We begin a general exploration of the phenomenology of TeV-scale extra-dimensional models with 
gravitational actions that contain higher curvature terms. In particular, we examine how the 
classic collider signatures of the models of Arkani-Hamed, Dimopoulos and Dvali (missing energy 
and new dimension-8 contact interactions) and of Randall and Sundrum (TeV-scale graviton Kaluza-Klein 
resonances) are altered by these modifications to the usual Einstein-Hilbert action. We find that not 
only are the detailed signatures for these  gravitationally induced processes altered but new 
contributions are found to arise due to the existence of additional scalar Kaluza-Klein states in 
the spectrum.
\end{abstract}



\renewcommand{\thefootnote}{\arabic{footnote}} \end{titlepage} 

%
%
%

\section{Introduction and Background}

The question as to why the Planck and electroweak scales differ by so many orders of 
magnitude remains mysterious. In recent years, attempts have been made to address this 
hierarchy issue within the context of theories with extra spatial dimensions that lower the 
effective scale of gravity to the TeV region. In both 
the models of Arkani-Hamed, Dimopoulos and Dvali (ADD){\cite {ADD}} and of Randall and Sundrum 
(RS){\cite {RS}}, new effects of gravitational origin are expected to occur near 
the TeV scale which should be observable at future colliders such as the LHC and ILC. Though 
these two models are very different in detail they do have some common features the most 
important of which are: ($i$) in their original versions they both assume that Standard Model matter 
is confined to a 4-dimensional brane; ($ii$) they both assume that D-dimensional gravity is 
described by the Einstein-Hilbert (EH) action plus a possible cosmological constant and 
($iii$) the background spaces are maximally symmetric and are either strictly 
flat, \ie , Minkowskian as in the 
ADD model with toroidally flat compactification, or being of constant curvature and is conformally 
flat, \ie, $AdS_5$ as in the RS model.{\footnote {In the original RS model the space is not truly one  
of constant curvature, \ie, the curvature invariant $R$ is not strictly co-ordinate independent 
as it also has delta-function singularities due to the presence of the TeV and Planck branes 
at the two $S^1/Z_2$ orbifold fixed-points. We will return to this issue later.}}  
How would the predictions of these two models be changed if we surrendered the assumption ($ii$), \ie,  that the 
EH action provided the full description of gravity 
and considered something more general? This is the discussion we would like to begin in this paper 
which will follow a phenomenological bottom-up approach. 

General Relativity (GR) as described by the EH action is considered to be an effective theory below 
the fundamental Planck scale, $M$. Thus, once energies approaching the scale $M$ begin to be probed one might 
expect to observe deviations from the expectations arising from the EH action. In the cases of both the 
ADD and RS models, future colliders will probe near or at their (effective) fundamental scales so that non-EH 
aspects of the true gravitational theory, whatever its form, should become apparent and be experimentally 
measured. Since the ultraviolet form of the true gravity theory is 
as of yet unknown one may hope to capture some of its deeper aspects by considering how the 
presence of new higher curvature (and higher derivative) invariants in the actions of the ADD and 
RS models can lead to variations in the well-known predictions of these theories. Many authors 
have considered the possibility of higher curvature invariants and how their existence 
would modify the predictions arising from the EH action within other contexts, \eg, the properties of  
black holes{\cite {BH,me}}, deviations in solar system tests of GR{\cite {solar}} and in cosmology{\cite {Carroll}} 
to possibly avoid the need for dark energy. Some analyses along these lines for the potential modifications 
of the collider predictions of both the ADD and RS 
models have already been performed{\cite {me,Demir:2005ps}}. In the present paper, we wish to both extend and 
generalize these results to get a feeling for the possible detailed variation of the various new gravitational 
phenomena as predicted by these classic models which 
will be potentially observable at future colliders. In particular we are interested in how 
the well known signatures of both the ADD and RS models are morphed if we keep the basic 
setups intact but modify the actions on which the corresponding equations of motion are based. 
A further generalization of such an analysis is possible if the original ADD/RS setups can now be simultaneously 
surrendered due to these modified actions resulting in entirely new setups with corresponding 
equations of motion; while 
this is an interesting possibility to consider it lies mostly beyond the scope of the present paper though it 
will be touched upon briefly in the discussion below. 

Of course a completely general study of how these possible modifications to the effective gravity action may 
morph TeV collider signatures is an obviously immense task and here we aim only at a 
first round analysis in the discussion that follows. The major signatures arising in both ADD and RS 
models originate from graviton exchange and the production of black holes; the ADD model also leads to 
missing energy signatures from graviton emission. Fortunately, apart from issues associated with black 
holes, since we are dealing with maximally symmetric spaces of globally constant curvature, the relevant graviton 
properties (couplings, 
wavefunctions and propagators) necessary to extract experimental signatures for either model 
can be obtained from the expansion of the rather general action considered below to quadratic order in the 
curvature. (This would no longer be true if we wanted to consider, \eg, the triple graviton coupling as then 
an expansion to third order in the curvature would be required.) 
This simplifying observation forms the basis of the analysis that follows and allows us to 
determine the relevant graviton properties in both ADD- and RS-like models for a wide class of effective 
actions. 

The general outline of our analysis is presented in Section 2 where our basic assumptions and notations 
are also given. In Section 3 we apply our analysis to the ADD model; we then apply it to the RS case in 
Section 4. Our summary and conclusions are given in Section 5.

\section{Analysis}

When going beyond the EH action there are many possibilities to consider especially when we are living in 
extra dimensions. In the literature various forms have been assumed for the potential structures of higher 
curvature and/or derivative invariants that may appear in the gravity action. The fairly general structure 
that we will assume for the D-dimensional action in the present analysis takes the form (with $D=4+n$ below):
\begin{equation}
S_g={{M^{D-2}}\over {2}} \int d^D x {\sqrt g} ~F(R,P,Q)\,,
\end{equation}
where $F$ is an arbitrary continuous, differentiable and generally mathematically well-behaved function; 
in particular we will assume that $F$ is non-singular when all of its arguments are zero.  
Here $R$ is the usual D-dimensional Ricci 
scalar while $P$ and $Q$ are quadratic invariants constructed from the curvature tensor 
$R_{ABCD}$: $P=R_{AB}R^{AB}$, with $R_{AB}$ being the Ricci tensor, while $Q=R_{ABCD}R^{ABCD}$. 
$M$ is the D-dimensional fundamental gravity scale which is $\sim$ TeV in ADD and $\sim \mpl$ in RS.
In the low energy, small curvature limit we expect $F \to R$ (plus a possible cosmological constant) and so the 
overall dimensionful factor in the expression above allows us to make direct contact with the EH action 
in this limit. This specific form for $F$, though not completely general, covers a wide array 
of possibilities and has been considered (sometimes only in $D=4$) in may different contexts for a multitude 
of purposes in the literature{\cite {BH,me,solar,Carroll}}. Many of the higher curvature models previous 
considered by other authors form subcases of this more general action. Generally we may think of $F$ as summing 
a number of corrections to the EH action which arise higher dimension invariants which are suppressed by appropriate 
powers of $M$. Thus we will only consider energy regimes where the collision energies are below the effective 
fundamental scale in what follows as is usually done when considering ADD and RS phenomenology. 

As is by now well-known{\cite {Accioly:2002tz,Schmidt:5}} the generalized gravity theories described by an 
action of the form $S_g$ can potentially have several serious problems. Since we will be dealing with ADD- and 
RS-type phenomenology we will be performing a perturbative analysis in the discussion that follows. Employing 
such an analysis one finds that, amongst other things, this action leads to equations of motion 
which are generally fourth order in the derivatives 
of the metric. In particular, in addition to the usual massless D-dimensional tensor graviton 
which results in, \eg,  the familiar 4-d graviton and graviscalar Kaluza-Klein (KK) tower excitations, there may 
also be present in the linearized D-dimensional theory additional {\it massive} scalar and tensor excitations. 
These fields will in 4-d have KK towers without massless modes and which can be ghostlike and/or tachyonic. (We 
can think of these new D-dimensional fields as having bulk masses which influence their corresponding 4-d 
KK tower masses.)  Furthermore, the equations of motion naturally involving higher derivatives of the fields can 
lead to problems with unitarity as well as non-canceling asymmetric pieces of the corresponding Einstein tensor. 
The new massive tensor excitations are potentially the most serious issue to deal with 
as they are ghost fields that must be eliminated from the perturbative 
spectrum (though they may help in dealing with the 
theory's renormalizability and bad high-energy behavior). It has been 
noted{\cite {Accioly:2002tz,Schmidt:5}} that we one can remove these states from the 
spectrum (\ie, by giving their bulk masses an infinite value) if a tuning occurs such that the quantities 
$P$ and $Q$ only appear in the special combination $Q-4P$ in the function $F$. How this tuning arises in the 
fundamental theory is unspecified. 
There has been some discussion in the literature, however, that these ghost states may not be 
as dangerous as one would imagine from lowest order perturbation theory{\cite {Antoniadis:1986tu}} so that we 
should perhaps keep an open mind about 
the possible forms for $F$. We will return to this point in what follows. 

Given a general action of the kind above there are several issues that one normally wants to 
address in order to extract information that can be compared with experimental data. From studies of both 
the ADD and RS models there are certain things we want to know, \eg, ($i$) the spectrum, 
wave functions, propagators and Standard Model (SM) matter couplings of the KK  
graviton (and other possible) excitations and ($ii$) the relationship between $M$, the volume 
of the compactified 
dimensions and the (reduced) 4-d Planck scale $\mpl$. {\footnote {As in the ADD and in the 
original RS models, in what follows we will assume that the SM matter fields are localized at 
a fixed value of the extra-dimensional co-ordinates.}} To obtain this information, as well as 
to make contact with several other analyses{\cite {me,Demir:2005ps}}, it is sufficient to expand the 
general action $S_g$ above around the maximally symmetric background metric to quadratic order in the curvature 
to obtain an effective action for the graviton (and like) excitations. At this level, one can extract 
the relevant 2-point functions as 
well as the differential equation for the KK wavefunctions which then yield the KK mass spectrum 
as well as the the desired graviton couplings to the localized SM fields. 
If, however, one wanted to probe, \eg, graviton 3- or 4-point functions then we would need to 
expand to at least cubic or quartic order in the curvature, respectively; these will not be of 
interest to us here but might be of interest in future experiments{\cite {Davoudiasl:2001uj}} which would 
tell us more about the underlying theory of gravity. Interestingly, if we allow for cubic or higher invariants 
in the original action in a maximally symmetric background, they will make no contributions to the quantities 
of interest to us here. Thus our analysis allows for the most general possible effective bulk local action for 
gravity. 

Once we make this expansion, there are various equivalent ways of expressing the resulting effective action, 
$S_{eff}$, depending upon the basis of invariants we choose to employ; the most obvious form is simply
\begin{equation}
S_{eff}^{(1)}={{M^{D-2}}\over {2}} \int d^D x {\sqrt g} ~\big[\Lambda+dR+aR^2+bP+cQ\big]\,,
\end{equation}
where $P,Q$ have been defined above. $\Lambda$ is an effective cosmological constant and 
$a-d$ are (in some cases dimensionful) constants all of which are functions of $F$ and its derivatives 
evaluated by employing the relevant background metric. To relate this back to the EH action in the limit of 
small curvature, one can think of the (necessarily positive) parameter $d$ as a `renormalization' 
of the fundamental mass scale $M$: $M\to M'=Md^{1/(D-2)}$. 

A second and perhaps more useful version of $S_{eff}$ is given by 
\begin{equation}
S_{eff}^{(2)}={{M^{D-2}}\over {2}} \int d^D x {\sqrt g} ~\big[\Lambda+dR+a'R^2+b'P+c'G\big]\,,
\end{equation}
where $G$ is the well-known Gauss-Bonnet(GB) invariant:
\begin{equation}
G=R^2-4R_{AB}R^{AB}+R_{ABCD}R^{ABCD}=R^2-4P+Q\,,
\end{equation}
The co-efficients $a',b'$ and $c'$ can be easily converted 
to $a-c$ above by some straightforward algebra: $a=a'+c',b=b'-4c'$ and $c=c'$. 
In D$\leq$4, the GB invariant is either a topological invariant or just a total derivative not contributing 
to the equations of motion, but this is no longer true for 
arbitrary values of $D>4$. The GB invariant is just (the quadratic) member of a general class of  
Lovelock invariants, constructed of various powers of the curvature tensor, which lead to 
special properties for the equations of motion{\cite {Lovelock}}. Generally the existence in the action 
of higher curvature terms, as discussed above, leads to higher order equations of motion that 
produce tachyonic and/or ghost excitations in the spectrum as well as potentially 
non-symmetric and/or non-conserved pieces of the corresponding Einstein tensor. 
Having an action consisting solely of Lovelock invariants avoids all of these potential 
difficulties as well as those associated with the massive tensor ghosts. The D-dimensional scalar 
excitation discussed above is also absent in this case. 
It is interesting to note that the GB term is the leading correction to the EH action
in perturbative string theory{\cite {Zwiebach,Mavromatos}}. Higher order Lovelock invariant may also be present in the 
action (when $D>6$) but these cannot be described by the function $F$ as employed here since they are 
constructed out of 
cubic or higher order combinations of the curvature tensor. The effect of the presence of general 
Lovelock invariants in the action of the ADD model has been discussed within the black hole 
context in Ref.{\cite {me}}. 

A further possibly very useful form for the quadratic action that is commonly used in the literature is 
\begin{equation}
S_{eff}^{(3)}={{M^{D-2}}\over {2}} \int d^D x {\sqrt g} ~\big[\Lambda+a_1R+a_2R^2+a_3C+a_4G\big]\,,
\end{equation}
where $C$ is the square of the Weyl tensor which can be expressed as{\cite {Maroto:1997aw}}: 
\begin{equation}
C=C_{ABCD}C^{ABCD}=Q-{{4P}\over {n+2}}+{{2R^2}\over {(n+3)(n+2)}}\,.
\end{equation}
where $n=D-4$ is the number of extra dimensions; the $a_i$ are linearly related to the coefficients 
$a-d$ above, \eg, $a_1=d$. This translation is simplified via the use of the identity{\cite {{Maroto:1997aw}}
\begin{equation}
aR^2+bP+cQ=-\Big[{{(n+2)b+4c}\over {4(n+1)}}\Big]G+\Big[{{4(n+3)a+(n+4)b+4c}\over{4(n+3)}}\Big]R^2
+{{n+2}\over {n+1}}\Big(c+{b\over {4}}\Big)C\,.
\end{equation}
Noting that the $a_{2-4}$ have dimensions of mass$^{-2}$ it is sometimes common in the literature 
to write 
\begin{eqnarray}
a_1/a_2&=&2(n+3)m_0^2\nonumber \\
a_1/a_3&=&-(n+2)m_2^2\,,
\end{eqnarray}
where $m_{0,2}$ are two mass parameters which are naturally 
$\sim M$ in the theory. One then finds that $m_{0,2}$ are 
directly related to the bulk masses of the D-dimensional massive scalar and tensor excitations 
discussed above. To avoid tachyons we apparently must demand that $m_{0,2}^2>0$ but even in such a case 
as we will see this the massive tensor field remains a ghost since the kinetic term for this field will have 
the wrong sign. 

Clearly all these forms for $S_{eff}$ are simply related. 
In what follows we will make use of all of the above forms of $S_{eff}$ and treat them 
interchangeably. 

Our first goal will be to explicitly calculate $S_{eff}$ in one of these `bases' from the more 
general $S_{g}$ in terms of $F$ and its derivatives.   
To begin we perform a Taylor series expansion of $F$ to quadratic order in all three arguments evaluating 
the result in the background metric, \eg,
\begin{equation}
F=F_0+(R-R_0)F_R+(P-P_0)F_P+(Q-Q_0)F_Q+{\rm {quadratic ~terms}}\,,
\end{equation}
where $F_0$ is a constant corresponding to the evaluation of $F$ itself in the fixed curvature background 
metric and $F_X=\partial F/\partial X|_0$; $X_0$ means that $X$ is to be evaluated 
in terms of the background metric which we here assume to be a space of constant curvature, \ie, 
a maximally symmetric space as is the case in both the ADD and RS models. Thus the quantities $R_0$, $P_0$, 
$Q_0$, $F_X$ and $F_{XY}$ are just numbers which depend on the explicit form of the metric and possibly 
the number of extra dimensions.{\footnote {Note that the quadratic terms in the Taylor expansion naturally 
involve factors of $P^2$ 
and $Q^2$ which are actually fourth order in the (dynamical) curvature; we drop these terms for 
consistency in the discussion which follows.}} In such a maximally symmetric space the Weyl tensor 
and corresponding invariant both vanish identically, \ie, $C_0=0$ and one further finds that 
\begin{eqnarray}
P_0&=&{{R_0^2}\over {n+4}}\nonumber \\
Q_0&=&{{2R_0^2}\over {(n+4)(n+3)}}\nonumber \\
G_0&=&{{(n+2)(n+1)R_0^2}\over {(n+4)(n+3)}}\,.
\end{eqnarray}
Note that in ADD $R_0=0$ since the metric is Minkowskian whereas in the $AdS_5$ RS bulk $R_0=-20k^2$ 
(away from the two branes) where the parameter $k$ originates from the usual RS metric
$ds^2=e^{-2k|y|}\eta_{\mu\nu}dx^\mu dx^\nu-dy^2$. 

Without making any further assumptions we obtain 
\begin{eqnarray}
\Lambda &=&F_0-R_0F_R+R_0^2(F_{RR}/2-\sigma F_P-\tau F_Q)+R_0^3(\sigma F_{PR}+\tau F_{QR})\nonumber \\
&+&R_0^4(2\sigma \tau F_{PQ}+\sigma^2 F_{PP}+\tau^2 F_{QQ})/2\nonumber \\
a_1&=&F_R-R_0F_{RR}-R_0^2(\sigma F_{RP}+\tau F_{RQ})\nonumber \\
a_2&=&\beta F_P+\epsilon F_Q+F_{RR}/2-R_0(\beta F_{RP}+\epsilon F_{RQ})-R_0^2[(\tau \beta+\epsilon 
\sigma)F_{PQ}+\sigma \beta F_{PP}+ \tau \epsilon F_{QQ}]\nonumber \\
a_3&=&\alpha F_P+\delta F_Q-R_0(\alpha F_{RP}+\delta F_{RQ})-R_0^2[(\tau \alpha+\delta 
\sigma)F_{PQ}+\sigma \alpha F_{PP}+ \tau \delta F_{QQ}]\nonumber \\
a_4&=&-\alpha F_P+\gamma F_Q+R_0(\alpha F_{RP}-\gamma F_{RQ})+R_0^2[(\tau \alpha-\gamma 
\sigma)F_{PQ}+\sigma \alpha F_{PP}- \tau \gamma F_{QQ}]\,,
\end{eqnarray}
where we have defined 
$\sigma=(n+4)^{-1}$, $\tau=2(n+4)^{-1}(n+3)^{-1}$, $\delta=4\alpha=(n+2)/(n+1)$, $4\beta=(n+4)/(n+3)$, 
$\gamma=-(n+1)^{-1}$ and $\epsilon=(n+3)^{-1}$. For the case of $n=0$ this reproduces 
the results give by, \eg, Navarro and Van Acoleyen in {\cite {solar}}. Note that if we make the 
assumption that $F$ is a function only of $R$ and the combination $Q-4P$ then $F_P=-4F_Q$ \etc ~and, also 
noting that $\delta-4\alpha$=0, we obtain $a_3=0$ so that the remaining expressions greatly simplify; we now obtain 
\begin{eqnarray}
\Lambda&=&F_0-R_0F_R+F_{RR}R_0^2/2+\lambda R_0^2[F_Q-R_0F_{QQ}+\lambda F_{QQ}R_0^2/2]\nonumber \\
a_1&=&F_R-R_0F_{RR}+\lambda R_0^2F_{RQ}\nonumber \\
a_2&=&-F_Q+F_{RR}/2+R_0F_{RQ}-\lambda R_0^2F_{QQ}\nonumber \\
a_4&=&-a_2+F_{RR}/2\,,
\end{eqnarray}
where the parameter $\lambda$ is given by 
\begin{equation}
\lambda={{2(2n+5)}\over {(n+3)(n+4)}}\,.
\end{equation}
Note that having $a_3=0$ implies that the Weyl term, $C$, in the effective action is absent in second order 
which is equivalent to taking $m_2\to \infty$ thus eliminating the massive tensor ghost issue. 
This field is now removed from the spectrum though the D-dimensional scalar remains in general.

\section{Application I:~ADD}

In this section we will apply the above analysis to the general ADD framework where we now require 
(since the space is flat) $R_0=F_0=0$ so that $\Lambda=0$ automatically. This significantly reduces 
the possible deviations from the classic ADD picture. In this specific case the general second order 
expansion of $F$ is rather simple and is given by  
\begin{equation}
F\to F_R R+\big[\beta F_P+\epsilon F_Q+F_{RR}/2\big]R^2+\alpha\big[F_P+4F_Q\big]C
+\big[-\alpha F_P+\gamma F_Q]G\,.
\end{equation}
Note that if we also demand that $F$ be a function only of $R$ and the combination $Q-4P$ in order to avoid 
the issue of the massive tensor ghost this expression simplifies even further to
\begin{equation}
F\to F_R R+\big[-F_Q+F_{RR}/2\big]R^2+F_Q G\,.
\end{equation}
Of course, there has been some discussion about other ways to circumvent this tensor ghost problem than by completely 
eliminating it from the perturbative spectrum. Since we are working only to lowest non-trivial order perhaps we 
should keep an open mind about the forms for $F$. Note that since $F_R$ will essentially rescale the overall 
mass factor in the action we must demand that $F_R>0$ to insure that the 
usual D-dimensional massless tensor gravitons not be ghost-like. 

How are the predictions of ADD modified by these additional curvature terms? {\footnote {This subject has 
already begun to be  
addressed by Demir and Tanyildizi in Ref.{\cite {{Demir:2005ps}} in the case where $F$ is only a function of R.}} 
The basic ADD picture leads to three essential predictions {\cite {JM}}: ($i$) the
emission of graviton KK states during the collision of SM particles
producing signatures with apparent missing energy{\cite {GRW,HLZ,PP}}; ($ii$)
the exchange of graviton KK excitations between SM brane fields leading to dimension-8 contact
interaction-like operators with distinctive spin-2 properties{\cite {GRW,HLZ,JLH};
($iii$) the production of black holes(BH) at colliders and in cosmic rays
with geometric cross sections, $\simeq \pi R_s^2$, with $R_s$ being the BH Schwarzschild radius, 
once collision energies greater than $\sim M$ are exceeded{\cite {Fisch,DL,GT,Kanti}}.

The production and properties of D-dimensional, TeV-scale BH in higher curvature theories has been 
partially explored within the context of Lovelock extended gravity{\cite {me}} though not yet so in 
the fully general quadratic gravity case described by the function $F$ 
considered here. Such a study, which would be very interesting, is 
far beyond the scope of the present analysis. However, it is interesting to make several observations: ($i$)
Consider the vacuum solution; if we expand 
the general action $F$ above to only quadratic order and {\it if} we also assume that all interesting 
solutions must satisfy $R_{AB}=0$ in the vacuum, {\it then} the only deviations from the conventional 
Schwarzschild form arise from the GB term in the action. This can be seen immediately by examining 
the equations of motion resulting from the general quadratic action, \eg, in Ref.{\cite {eom}}. This result 
does not remain valid when $F$ is treated exactly. ($ii$) If $F=F(R)$ only and is treated exactly without expansion 
then the equations of motion allow for the conventional external BH result  
with $R_{AB}=0$ and will appear as an ordinary D-dimensional 
Schwarzschild solution. This is {\it not}, however, the most general solution as can be see by considering the simple 
case of $F=R+\beta R^2/M^2$ with $\beta$ a dimensionless parameter. Here there is also exists a solution with 
$R\sim -M^2/\beta$, which is a constant corresponding to deSitter or anti-deSitter space depending on the sign 
of $\beta${\footnote {See, however, Ref.{\cite {Hindawi:1995cu}}}}. ($iii$) If $F=R+\alpha G/M^2$, with $\alpha$ 
being a constant, then the general BH solution has neither $R$ nor $G$ equal to zero as is well-known from the 
exact solution{\cite {{BH,Wheeler,Mignemi:1991wa}}. A more detailed study of these possibilities would be worthwhile.  

Let us now consider the situation of graviton exchange where it is well-known{\cite {GRW,HLZ,JLH}} that ADD leads 
to new dimension-8 contact interactions. To obtain the analogous quantities here we must expand the integrand 
of the action, \ie, the Lagrangian $\sqrt g F$, to second order in the fluctuations, $h_{AB}$, around the 
flat background metric:
\begin{equation}
{\cal L}_2={1\over {2}}h^{AB}O_{ABCD}h^{CD}\,.
\end{equation}
Here we have expressed the original metric as 
\begin{equation}
g_{AB}=g_{AB}^0+2h_{AB}/M^{1+n/2}\,,
\end{equation}
with $g_{AB}^0$ being the background metric, which is identified in the ADD model as the Minkowski metric 
$\eta_{AB}$. The propagator is then just the 
inverse of the operator $O$; once the propagator is known we sandwich it between two 4-d localized (at the 
origin of the extra-dimensional co-ordinates) and conserved SM  
stress-energy sources, $T_{\mu\nu}$, to find the relevant scattering amplitude; we must remember to later 
KK decompose the various towers. Fortunately, much of this work has been done for us by Accioly, Azeredo and 
Mukai{\cite {Accioly:2002tz}} 
from which, with some modifications, we obtain the expression for the D-dimensional `graviton' exchange amplitude 
(before performing the KK sums)
\begin{equation}
{\cal A} \sim {{T_{\mu\nu}T^{\mu\nu}-T^2/(n+2)}\over {k^2-m_n^2}}-{{T_{\mu\nu}T^{\mu\nu}-T^2/(n+3)}
\over {k^2-(m_2^2+m_n^2)}}+{{T^2}\over {(n+2)(n+3)[k^2-(m_S^2+m_n^2)]}}\,,
\end{equation}
where $m_n^2=\sum_i n_i^2/R_c^2$ are the familiar (unaltered in terms of the compactification radius $R_c$) 
flat space KK masses. The $n_i$ label the various KK levels; 
$T=\eta^{\mu\nu}T_{\mu\nu}$ is the 4-d trace of the SM source stress-energy tensor and $m_{S,2}$ are just 
\begin{eqnarray}
m_S^2&=&{{(n+2)m_0^2}\over {2}}={(n+2){F_R}\over {4(n+3)(\beta F_P+\epsilon F_Q+F_{RR}/2)}}\nonumber \\
m_2^2&=&{{-F_R}\over {(n+2)(F_P+4F_Q)}}\,, 
\end{eqnarray}
as described above. $R_c$ is the compactification radius which sets the KK mass scale; here we have assumed a
common value for this quantity for all $n$ extra dimensions so that the volume of the compactified space is 
just $(2\pi R_c)^n$. In the expression above the 
first term in the amplitude is the usual one encountered in the ADD model which results from the D-dimensional 
EH action and combines the contributions of the 4-d spin-2 graviton and spin-0 graviscalar KK towers.
The second and third terms correspond to the new D-dimensional massive ghost tensor and scalar contributions, 
respectively. The difference in the factors of $n+2$ versus $n+3$ in the first two terms arises from the 
existence of a 5-d bulk mass for the tensor ghost field. It is interesting to note that the full amplitude is 
very well behaved at large $k^2$ (in fact, going as $\sim 1/k^4$) due to the detailed cancellations between the 
various terms. 

Here 
$m_{S,2}$ represent the bulk mass terms of the new fields which enter into the KK tower masses of the 
scalars (spin-0) and tensors (spin-2 and spin-0), 
respectively; here we see the effect of the tensor ghost KK tower exchange explicitly. 
From this point of view it 
appears that the only way to remove this ghost tower is to take the bulk mass $m_2^2 \to \infty$ 
implying that $F$ is solely a function of $R$ and the combination $Q-4P$, which we will assume from now on in our 
ADD discussion. 
Note that the existence of a GB term in the action will not yield a contribution to $m_{S,2}$. 
Since, as discussed above $F_R>0$ is 
already required, tachyonic KK scalars are avoided when the denominator in the expression for $m_S^2$  
above is positive; when $F$ is assumed to be a function only of $R$ and $Q-4P$, then this denominator simplifies 
to $F_{RR}/2-F_Q$.  In the limit where $F=F(R)$ alone, and accounting for a sign factor in the 
definition of the above actions, our result for the squared 
scalar mass, $m_S^2$, agrees with that obtained by Demir and 
Tanyildizi{\cite {{Demir:2005ps}}. As shown by these authors, the effect of the new scalar tower exchange is generally 
rather suppressed in comparison to the more familiar graviton exchange since the ratio $T^2/T_{\mu\nu}T^{\mu\nu}$ is 
small for most SM particle sources at TeV colliders. For example, for the process $WW\to WW$ this ratio is of order 
$\sim (M_W^2/s)^2$. The corresponding ratio of the KK summed scalar to graviton exchange amplitudes 
is somewhat further reduced by ($i$) the existence of the finite bulk scalar mass which implies  
that there are no light scalar KK exchanges with masses below $m_S$ and ($ii$) the $n$-dependent numerical factor 
in the denominator of the scalar amplitude. Naturalness suggests that $m_S \sim M \sim$ TeV 
or larger unless the parameters of $F$ are somehow fine-tuned. For example, if $F=R+\beta R^2/M^2$, then $m_S^2 
=(n+2)M^2/(4\beta (n+3)) \sim M^2$ for all $n$ {\it if} $\beta$ is not too far from 
O(1). Interestingly we see here that as $\beta \to 0$ we recover the 
usual EH expectation as then $m_S \to \infty$. Thus we find that for many practical 
purposes the structure of the usual ADD results for graviton exchange are not qualitatively modified when the action 
is generalized to the form considered here. However, with the existence of these additional scalars being a 
hallmark of the extended action, it behooves us to find a way to isolate their effects experimentally. 

In expressions for graviton exchange only the 
combination $M^{n+2}F_R$ will now appear. In the amplitude this will lead to a modification of the pure 
`graviton' exchange cross section 
expectations by a factor of $F_R^{-1}$, which is likely to be of $O(1)$, {\it provided} $M$ is considered 
to be held fixed. When the graviton tower interference term with the SM dominates, the effect in the gravitational 
part of the cross section will scale as  
$F_R^{-1}$. Given the previous results of Demir and Tanyildizi{\cite {{Demir:2005ps}}, this is not surprising. 

We further note that since $m_S$ is $\sim$ TeV or larger it has no effect 
on laboratory measurements of the strength of the gravitational interaction in the micron range when $n=2$. 

Before closing this part of the discussion we would like to remind the reader that it was pointed out 
long ago{\cite {Schmidt:5,Maeda:1988ab}}  
that we can take any action of the form $F=F(R)$ and map it over to the EH action coupled to an minimally 
coupled real scalar field with a rather complicated potential $V$, depending exponentially on the scalar field.  
This can be done via a special conformal transformation  
\begin{equation}
\hat g_{AB}=|F_R|^{2/(n+2)}g_{AB}\,. 
\end{equation}
Going from the original (Jordan) to the new (Einstein) frame one explicitly sees the existence 
of the new scalar degree of 
freedom{\footnote {It is even possible to make a generalized version of this transformation for more complicated 
forms of the action{\cite {Jakubiec:1988ef}}}}. The mass of this 
scalar field is exactly that of the field $S$ above and can be gotten directly from the canonically normalized 
potential $V$ in the usual manner, \ie, using $m_S^2=\partial^2 V/\partial S^2$. This is a very powerful tool as it 
allows us to extend our previous flat space result for $m_S$ to the much more general case 
where the space has constant curvature. For example, if $F=R+\beta R^2/M^2$, we find that the value of $m_S^2$ is the 
same as discussed above, \ie, $m_S^2=(n+2)M^2/(4\beta (n+3))$, in a space with constant curvature. 
This will be an important result that we will employ when we discuss the case of the RS setup.

We now turn to the emission of gravitons in SM particle collisions. 
Since the compactifying space is flat in the ADD case the 
normalizations of the graviton (and scalar) wavefunctions which control their couplings are 
unaltered by the existence of the quadratic curvature terms but the relationship between $M$ and $\mpl$ 
is modified. This was briefly mentioned above where we saw that in the small curvature limit the 
parameter $d$ essentially renormalizes the fundamental scale. 
To see this in the present case it is sufficient to examine the tensor/spin-2 kinetic part of the 4-d effective 
Lagrangian to second order in $h_{\mu\nu}$ (which has not yet been KK-expanded) in the familiar 
transverse traceless gauge, \ie, $\partial_\mu h^{\mu\nu}=0$, $h_\mu^\mu=0$; one obtains{\cite {Accioly:2002tz}}
\begin{equation}
{\cal L}_g=-{d\over {2}}h^{\mu\nu}\Box h_{\mu\nu}+{1\over {2}}\Big({b\over {4}}+c\Big)\Box h^{\mu\nu}
\Box h_{\mu\nu}\,,
\end{equation}
where here $\Box=\eta^{AB}\partial_A \partial_B$ and $b,c,d$ are defined above{\footnote {Note that 
as usual Greek indices only run over 4-d.}}. When we assume that $F$ is only a function of $R$ and the combination 
$Q-4P$ then the second term in ${\cal L}_g$ vanishes and we recover the familiar result of the standard 
EH scenario apart from the overall factor of $d$. Hence, to recover the conventional 4-d EH action when 
inserting the usual (extra dimensionally) flat zero mode graviton wavefunction into ${\cal L}_g$ the  
ADD relationship must be modified, as hinted above, to  
\begin{equation}
\mpl^2=V_nM^{n+2}F_R\,,
\end{equation}
where $\mpl$ is the 4-d reduced Planck scale and $V_n$ is the volume of the compactified space. Of 
course, $d=F_R$ is just unity in the standard ADD model which employs the EH action. Since the lightest of the 
KK scalars has a mass which is naturally on the order of a TeV and has rather weak couplings to SM fields 
these particles will not play much of an important role in missing energy processes.  
{\it If} the cross section for graviton production, \ie, missing energy, is 
expressed in terms of the original $M$ with other parameters held fixed, then the presence of $F_R$ 
leads to a modification of the production cross section by a factor of $1/F_R$. However, as $M$ is not likely 
to remain a direct observable (only the product $M^{n+2}d$ is) there may be no way to experimentally disentangle  
this effect. Furthermore, for any given $F_R$, since $\mpl$ is numerically fixed and $M$ is an input parameter 
the resulting derived value of $R_c$ which sets the scale for the masses of the KK states is altered. 

We thus conclude that if we assume that $F$ is a function of only of $R$ and the combination $Q-4P$ then the 
classic predictions ($i$) and ($ii$) of the ADD model will be qualitatively unaffected by going to the   
more general action considered here {\it except} for possible overall scalings by inverse powers of $F_R$ 
when the parameter $M$ is held fixed: graviton emission rates scale like $1/F_R$ while graviton exchange 
cross sections scale as $1/F_R$ or $1/F_R^2$ depending on the presence of important SM contributions to the 
relevant process. Furthermore, given Eq.(22) and fixed values of $M$ and $F_R$, the KK masses, being proportional 
to $1/R$, will scale as $m_{KK}\to m_{KK}F_R^{1/n}$; such a mass shift can be quite sizable for reasonable 
values of $F_R$.

\section{Application II:~RS}

The predictions of the classic RS model are the existence of TeV scale graviton resonances with fixed 
weak scale masses and couplings to the SM fields{\cite {Davoudiasl:1999jd}}, the existence of a weak scale 
radion excitation{\cite {Giudice:2000av}}, as well as the production of $AdS_5$ BH. In what 
follows we will be specifically interested in the nature of the KK gravitons so it is again sufficient to
examine the quadratically expanded action. The classic RS model is not generally strictly 
consistent with the assumed 
form of either the original action $S_g$ or its quadratically expanded form $S_{eff}$. As is well-known, and 
as mentioned above, the equations of motion that follow from $S_g$ and $S_{eff}$ will generally be 
fourth order in the derivatives of the metric. In the usual 5-d RS model, one solves the Einstein 
equations of the form 
\begin{equation}
G_{AB}={{T_{AB}}\over {M^3}}\,,
\end{equation}
where $G_{AB}$ is the Einstein tensor arising from the EH action involving no more than two derivatives 
of the metric. The problem is that RS completely specifies $T_{AB}$: a cosmological constant in the 5-d 
bulk plus two $\delta$-function sources at the orbifold locations of the TeV and Planck branes. SM 
matter confined to the TeV brane is supposed to not be a large contributor to the stress-energy. To obtain 
this result the standard RS 
metric takes the form discussed above: $ds^2=e^{-2k|y|}\eta_{\mu\nu}dx^\mu dx^\nu-dy^2$ 
with the linear exponential warp factor leading to the bulk $AdS_5$ and the two field derivatives acting on 
the absolute value leading to the brane $\delta$-functions. (This is related to the comment above that $R$ is 
not truly constant in RS and has brane $\delta$-function singularities. Recall that these $\delta$-functions 
are the results of assuming infinitely thin branes.) 
If an identical metric is assumed in our more general case we still can obtain 
$AdS_5$ but the fourth order equations would lead to the more singular derivatives of $\delta$-functions 
at the brane locations that are not canceled by any source terms. 
This amongst other reasons is what led Kim, Kyae and Lee{\cite {KKL}} to 
consider only GB extensions of the EH action in RS since it is the only extension which uniquely 
produces Einstein equations of second 
order in the derivatives. Thus if we keep the classic picture, an analysis of RS given our assumed effective 
action expanded around a background of constant curvature is not relevant. (A possible way of 
dealing with these derivatives of $\delta$-functions arising from orbifold singularities in higher dimensional 
effective field theories has been discussed in Ref.{\cite {delAguila:2006kj}}. Implementing our scheme employing 
such techniques is, however, beyond the scope of the present paper.)

To avoid these issues for now we simplify our discussion of this problem (and to convince ourselves that an 
RS-like solution is possible in this framework)  
we consider a singularity-free, `softened' version of RS where the orbifolded bulk space with branes is replaced 
by an interval, as has been suggested for other reasons{\cite {Csaki:2003dt}}, with SM matter placed 
at one end point possessing an ignorable amount of stress-energy. With a 
cosmological constant on the interval we can recover the background $AdS_5$ bulk; in addition 
by removing the absolute value sign of the co-ordinate $y$ in the metric above 
we expunge the $\delta$-functions as well as the possibility of any 
of their higher derivatives appearing in the equations of motion. The boundary conditions at the end points 
for the graviton KK states can then be freely chosen to be the same as that of the original RS model.   
This space is truly one of constant curvature and the general analysis we have presented above will now be 
applicable to this `softened' RS on an interval. 

It is easy to verify that the form of the equations of motion{\cite {eom}} in this case (recalling that we are 
only searching for solutions with maximally-symmetric, constant $R$ backgrounds) are given by: 
\begin{equation}
F_RR_{AB}-{1\over {2}} g_{AB}F+2F_PR^N_A R_{NB}+2F_Q R_{MNSA} R^{MNS}_{~~~~~B}={{T_{AB}}\over {M^3}}\,,
\end{equation}
and that if we take stress-energy tensor in the 5d bulk to be of the usual RS form  
\begin{equation}
T_{AB}=-\Lambda g_{AB}\,,
\end{equation}
with $\Lambda>0$, 
then indeed a space of constant curvature, \ie $AdS_5$, {\it can} be an allowed solution. Taking the trace of 
the equations of motion above, evaluating it in the constant curvature bulk and relating the values of 
$P_0,Q_0$ to $R_0^2$ as before (recalling that here $R_0=-20k^2$ using the softened metric) results in the 
constraint equation   
\begin{equation}
{2\over {5}} R_0^2\big[F_P+{1\over {2}}F_Q\big]+F_R R_0-{5\over {2}}F_0={T\over {M^3}}=-{{5\Lambda}\over 
{M^3}}\,,
\end{equation}
where here $T=g^{AB}T_{AB}$. 
It is interesting to note that if we {\it assume} that $R=$constant then this constraint equation automatically 
implies that $T=$constant as well; but this does {\it not necessarily} further require that all of the $T_{AB}$ are 
constants as we will see below. 
When $F$ is only a function of $R$ and the combination $Q-4P$, this constraint equation simplifies to 
\begin{equation}
-{7\over {5}} R_0^2F_Q+F_R R_0-{5\over {2}}F_0=-{{5\Lambda}\over {M^3}}\,;
\end{equation}
while in the specific RS background case this explicitly becomes 
\begin{equation}
224k^4F_Q+8k^2F_R +F_0={{2\Lambda}\over {M^3}}\,.
\end{equation}
It is important to recall that $F$ itself can be a complicated function of $k$ so that this equation can 
be quite nontrivial. 
For the EH action limit this yields the usual relation that $\Lambda=-6k^2M^3$; here it in general provides 
an additional constraint on the 
allowed forms of the function $F$ since we are requiring $R_0$ to be both real and negative. Given a specific 
function $F$ for which a solution exists, this equation directly relates $\Lambda$ and $k^2$ though the 
solution may not be unique. For example, if we assume for purposes of demonstration the simple case of  
\begin{equation}
F=R+{\beta\over {M^2}} R^2\,,
\end{equation}
as employed above, then there are two branches of solutions for $k^2(\Lambda)$:
\begin{equation}
k^2={{3M^2}\over {40\beta}}\Big[1\pm \Big(1+{{40\Lambda}\over {9M^5}}\beta\Big)^{1/2}\Big]\,,
\end{equation}
one of which (the negative root) goes over to the usual EH result as the parameter $\beta \to 0$.

Allowing for the possibility of a RS-like solution with a softened metric 
it is interesting to think briefly about the previously analyzed effects of the GB term in the RS scenario. 
This analysis was originally performed for the classic RS{\cite {me}} setup 
which employed the standard form of the RS metric; that result would now be modified by the  
changes in the model assumptions, \ie, moving to an interval and removing the $\delta$-function sources at the 
end points. The previous analysis of BH in RS with the added GB term would not be significantly affected if this 
transition were made. However, the properties and 
spectrum of the graviton KK states certainly would be influenced since the $\delta$-function terms are now absent. 
The equation governing the masses and wavefunction of the graviton KK states for the present interval case 
can be obtained by expanding the 
equations of motion as before. Since we are here only interested in the tensor modes associated 
with the usual gravitons, we can employ the expansion 
\begin{equation}
g_{\mu\nu}=e^{-2ky}\big(\eta_{\mu\nu}+\kappa h_{\mu\nu}\big)\,,
\end{equation}
where $\kappa=2M^{-3/2}$. Applying the usual RS boundary conditions on the interval the most significant changes 
from the classic RS can be read off from Eqs.(15)-(28) in Ref.{\cite {me}} by setting the parameter $\Omega=0$ 
in appropriate places. At the end of the day we find that 
the only apparent difference from the classic EH based RS model would be a shift 
in the relationship between the fundamental scale and $\mpl$--remarkably similar to what we saw  
for the ADD model above. In the language employed in Ref.{\cite {me}} we would now obtain 
\begin{equation}
\mpl^2={{M^3}\over {k}}\Big[1-4\alpha {{k^2}\over {M^2}}\Big]\,,
\end{equation}
where $\alpha/M^2$ is the coefficient of the GB term in the action. Otherwise the masses as well as the 
couplings of all of the KK gravitons to localized SM matter would be {\it identical} to those of the original RS 
model expressed in terms of the derived parameter $k$. 
The explicit coupling and spectrum changes found in Ref{\cite {me}} for the graviton KK states in the presence of 
the GB term in the action were all found to due to the brane $\delta$-function singularities. 

How would these graviton KK results obtained in the GB extended action generalize to the case of $S_{eff}$ above? 
Here we choose to begin our analysis with $S_{eff}^{(3)}$,  setting $a_3=0$ from the beginning to avoid potential 
ghost fields, then taking $D=5$ and using the same curvature expansion as above. In order to  
make a connection with the previous discussion, the existing RS literature and to directly 
compare with the GB case, however, we massage 
our notation slightly and rewrite $S_{eff}^{(3)}$ in the following form: 
\begin{equation}
S_{eff}^{(3)}=\int d^5 x {\sqrt g} ~\big[-\Lambda_b+a_1{{M^3}\over {2}}R+{{\alpha M}\over {2}}G+
{{\beta M}\over {2}}R^2 \big]\,,
\end{equation}
where the parameters $a_1, \alpha$ and $\beta$ are dimensionless; the action employed in Ref.{\cite {me}} 
is now directly recovered by taking the $a_1 \to 1$ and $\beta \to 0$ limits. 
It is important at this point to recall that to 
obtain the linearized graviton equations of motion it is sufficient to employ $S_{eff}$ while the complete 
$S_g$ needs to be examined in order to demonstrate the existence of the required $AdS_5$ solution. The equations of 
motion resulting from $S_{eff}^{(3)}$ are given by{\cite {eom}}
\begin{eqnarray}
-{\Lambda_b\over {M^3}} g_{AB}&=& a_1(R_{AB}-{1\over {2}}g_{AB}R)+{{2\beta}\over {M^2}}R(R_{AB}-{1\over {4}}g_{AB}R)
+{{2\beta}\over {M^2}}(g_{AB}\Box-\nabla_A \nabla_B)R \nonumber \\ 
&+& {{2\alpha}\over {M^2}}\Big[RR_{AB}-2R_{ASBP}R^{SP}+R_{ASPT}R^{SPT}_{~~~~B}-2R_{AS}R^S_B-{1\over {4}}g_{AB}G\Big]\,. 
\end{eqnarray}
Here $\nabla_A$ is the covariant derivative operator and here $\Box =g^{AB}\nabla_A \nabla_B$. First we look at 
the $A,B=5$ component of this equation, remembering that for the moment we will only be interested in the tensor 
excitations corresponding to the KK gravitons which are massless in 5-d. In the usually chosen gauge, $R$ is still 
a constant to linear order so we arrive at a consistency condition  
\begin{equation}
a_1-{{2k^2}\over {M^2}}\alpha-{{20k^2}\over {3M^2}}\beta=-{\Lambda_b\over {6k^2M^3}}\,. 
\end{equation}
Note that this reduces to the previously obtained purely quadratic GB extended RS result{\cite {me}} when $a_1=1,
\beta=0$. In the more general case, this expression is not overly useful given the exact result in Eq.(27). 

Turning now to the $A,B=\mu,\nu$ terms which contain the 4-d graviton tensor excitation, we linearize employing the 
previously mentioned transverse, traceless gauge with constant $R$. This gives the standard equation of motion 
for the RS graviton found long ago{\cite {Davoudiasl:1999jd}} though scaled 
by an overall factor. Employing the standard KK decomposition
\begin{equation}
h_{\mu\nu}(x,y)=\sum_n h^{(n)}_{\mu\nu}(x)\chi_n(y)\,, 
\end{equation}
and recalling that $\eta^{\alpha \beta} \partial_\alpha \partial_\beta  h_{\mu\nu}^{(n)}=-m_n^2 h_{\mu\nu}^{(n)}$, 
the $\chi_n$ are seen to satisfy
\begin{equation}
H\Big[\partial_y^2-4k\partial_y+m_n^2e^{2ky}\Big]\chi_n=0\,. 
\end{equation}
The overall factor $H$ is given by 
\begin{equation}
H=a_1-{{4k^2}\over {M^2}}\alpha-{{40k^2}\over {M^2}}\beta\,, 
\end{equation}
or, more explicitly in the RS case, 
\begin{equation}
H(k)=F_R+36k^2F_Q+1000k^4F_{RQ}+10080k^6F_{QQ}\,. 
\end{equation}
(Again we recall that $F$ itself can be a function of $k$.)
This leads to a rescaling of the usual RS relationship 
\begin{equation}
H(k){{M^3}\over {k}}=\mpl^2\,, 
\end{equation}
via the renormalization of the zero mode (\ie, massless graviton) wavefunction, thus generalizing Eq.(31). Of course, 
$H>0$ is required to avoid ghost states among the usual gravitons KKs.  This result reduces to that 
previously obtained in the RS case with just the added GB term{\cite {KKL,me}} once boundary effects are neglected. 

From this analysis we see immediately that the masses of the KK gravitons are identical to those obtained in 
the original RS model, {\it provided} we use the {\it same} value of the parameter $k$, as we might have expected. Here 
we are faced with the question of just what are the independent parameters. $k$ is clearly a {\it derived} parameter 
that is obtained by simultaneously solving Eqs.(27) and (40) 
for any given model. In that sense, the KK graviton spectrum would just be rescaled in comparison 
to the usual expectations given the same input value of $M$.  As we have just seen, and as in the 
ADD case, the effect of a factor like $H$ on the KK graviton couplings to 4-d SM matter depends upon which 
model parameters are assumed to be held fixed. At the very least, up to an overall constant, these KK graviton 
couplings are identical to those of the standard RS model. 

As an example of a simple model where the shifts in the KK spectrum can be calculated analytically 
consider substituting for the integrand of the conventional RS action, ${{M^3}\over {2}}R-\Lambda_0$, the 
simple higher curvature action  ${{M^3}\over {2}}(R+{\beta \over {M^2}} R^2)-\Lambda$ 
as was considered above. Let $k_0$ be the values obtained for the parameter $k$ in the usual RS model, \ie, 
$k_0^2=-\Lambda_0/6M^3$. Keeping 
the warp factor fixed we can use the equations above to calculate the value of, \eg, the mass of the 
first graviton KK state 
in both the standard RS model, $m_0$, and in the current model with an augmented action, $m$. 
Using Eqs.(27) and (40) this ratio can be calculated 
analytically in the present case as a function of $\beta$ and $c=k_0/\mpl$; we obtain
\begin{equation}
{m\over {m_0}}= (80\beta c^{4/3})^{-1}\big[-1+(1+160\beta c^{4/3})^{1/2}\Big]\,.
\end{equation}
The result of this calculation is shown in Fig.1 for a wide range of model parameters. In this example 
we see that the size of the possible shift in the mass spectrum can be quite large assuming a fixed value of $M$.

\begin{figure}[htbp]
\centerline{
\includegraphics[width=8.5cm,angle=90]{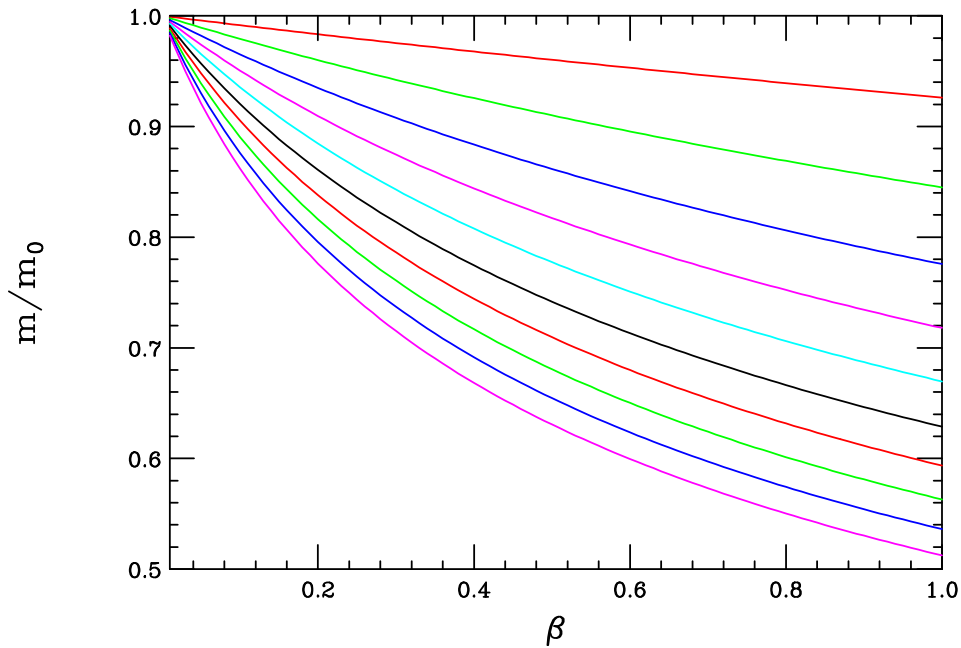}}
\vspace*{0.1cm}
\caption{KK mass shifts as a function of the parameter $\beta$ for different values of the ratio $c=k_0/\mpl$; 
here this ratio takes the values 0.01 to 0.10, corresponding to the curves from top to bottom, in steps of 0.01.}
\label{fig0}
\end{figure}
\begin{figure}[htbp]
\centerline{
\includegraphics[width=8.5cm,angle=90]{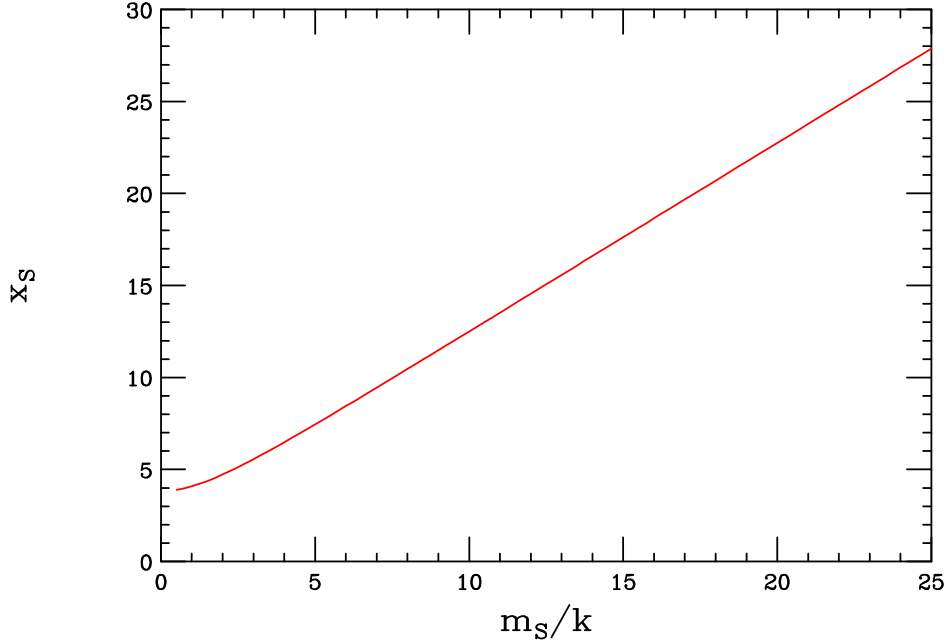}}
\vspace*{0.1cm}
\caption{Root for the determination of mass of the lightest KK state corresponding to the new RS scalar as a 
function of the scaled bulk mass. As a point of comparison the first KK graviton has a root of $\simeq$ 3.83.}
\label{fig1}
\end{figure}

So far we have only considered the 4-d graviton, spin-2 excitations. 
It is important to remember that our softened RS model now has an additional massive scalar in the 5-d spectrum 
with a large bulk mass, $m_S$, and that no massless scalar zero mode will exist. Since the bulk scalar mass is 
naturally of order $k$ the KK spectrum of the corresponding tower will begin with a KK scalar state whose mass is  
qualitatively comparable to that of the first graviton excitation. This bulk mass is explicitly 
calculable from the expansion 
of the full action to quadratic order, $S_{eff}^{(3)}$, by going to the Einstein frame since we know that the GB 
term does not contribute to this parameter. In that case, using the results from the previous section we find that  
\begin{equation}
m_S^2={{3a_1\over {16\beta}}}M^2\,, 
\end{equation}
or, in terms of the original parameters of the action, evaluated in the RS background: 
\begin{equation}
m_S^2={3\over {8}}~{{F_R+20k^2F_{RR}+280k^4F_{RQ}}\over {F_{RR}-2F_Q-40k^2F_{RQ}-560k^4F_{QQ}}}\,. 
\end{equation}
Note that $a_1,\beta>0$ is required to avoid the scalar tachyons and graviton ghosts, consistent with our above 
analysis. Note further that this reproduces the results of Eq.(19) in the flat space, $k \to 0$, limit.  

Given any $F$ the scalar bulk mass is known and we can determine the mass(es) of the lightest KK scalar state(s) 
by following the standard RS 
manipulations{\cite {Goldberger:1999uk}}. These masses are essentially given by the first roots of the equation 
\begin{equation}
(2-\nu)J_\nu(x_S)+x_SJ_{\nu-1}(x_S)=0\,, 
\end{equation}
where $\nu^2=4+m_S^2/k^2$ and $J$ is the usual Bessel function. The solution for the first KK state  
is provided by Fig.2; as stated above there are no massless modes. The lightest scalar mass is then 
$x_S ke^{-\pi kr_c}$. Here we observe that the mass of 
the first scalar KK scales almost linearly with the bulk mass when $m_S$ gets large. Note that for $\beta/a_1=1$ and 
a typical value{\cite {Davoudiasl:1999jd}} of $k/M=0.05$, we then 
find  $m_S/k \simeq 8.7$ implying $x_S \simeq 11$ from Fig.1; 
this is about 3 times larger than the root for the usual lightest massive KK graviton, $\simeq 3.83$. Thus we see 
that unless $\beta/a_1$ takes on large values the first scalar KK state is always rather heavy. As is well-known, the 
$x_S$ values for the more massive KK scalar states will be somewhat larger:
approximately given by $x_S \to x_S+(p-1)\pi/2$ where $p$ labels the KK level. 
Since these scalars will couple to the trace of the stress-energy tensor for the 4-d SM fields they will 
interact far more weakly than do the graviton KK states unless this is at least partially offset by ratios of 5-d 
wavefunction factors. A quick estimate of such factors, however, indicates that, if anything, these wave 
function ratios lead to a further suppression of the scalar couplings relative to those of the KK gravitons 
by $\simeq[12(1+(m_S^2/kx_S)^2)]^{-1/2}\simeq 1/4$ as shown in Fig.3. 
This overall picture of the scalar sector is qualitatively very similar to that of the existence of a very heavy 
tower of RS radions{\cite {{Giudice:2000av}} or a tower of KK Higgs bosons as in the case of Universal Warped Extra 
Dimensions{\cite {Davoudiasl:2005uu}}. 

In the analysis as presented here we have ignored 
the possibility that the new scalar KK states may mix with the (usually eaten) RS graviscalars through cross-talk 
in the equations of motion, \ie, we have assumed that the 5-d tensor 
and scalar KK decompositions can be performed independently, and this is something which needs further exploration. 
A fully detailed analysis of the such possibilities is, however, beyond the scope of the present paper.  

\begin{figure}[htbp]
\centerline{
\includegraphics[width=8.5cm,angle=90]{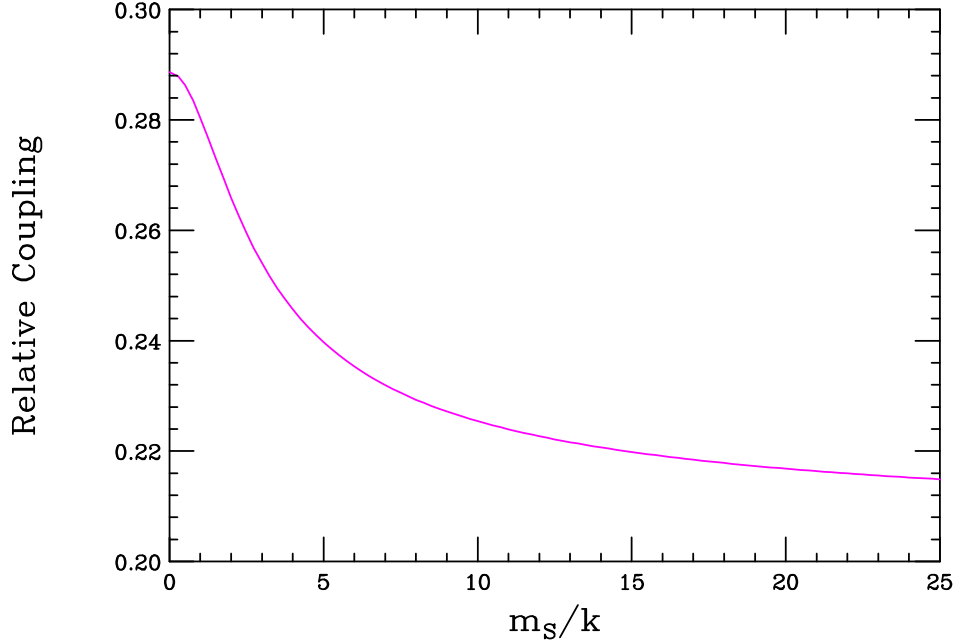}}
\vspace*{0.1cm}
\caption{Estimate of the lightest scalar KK coupling strength relative to that of the lightest KK graviton. Recall 
that graviton KK states couple to $T_{\mu\nu}$ whereas the new scalar KKs couple to its trace, $T$.}
\label{fig2}
\end{figure}

It is perhaps interesting to ask whether the usual $R$=constant ($R_0$) solution considered here necessitates 
the metric 
and matter distribution of the conventional RS model employed above without the further assumption of a maximally 
symmetric space. To analyze a simple and more easily tractable 
situation let us consider the more general warped metric $ds^2=e^{-2A(ky)}\eta_{\mu\nu}dx^\mu dx^\nu-dy^2$ and assume 
that $F=F(R)$ only. The first question to address is what is the most general form of the function $A$; to deal 
with this issue we note that the Ricci scalar arising from this metric is in general given by $R=R_0=-20A'^2+8A''$. 
In the `soft' version of the RS model on the interval defined above one had $A'=k$ and $A''=0$ so that $R_0=-20k^2$ 
as usual. Here, this curvature condition provides a differential equation for the function $A$; by solving this 
equation we arrive at the result  $A=c_1-{2\over {5}}{\rm{ln ~cosh}}[5a(ky+c_2)/2]$, where the $c_i$ are integration 
constants and $a=\sqrt{-R_0/20k^2}$. Choosing the $c_i$ appropriately and rescaling $y$, we can 
rewrite this in a more familiar form as $e^A=e^{aky}[1+\sigma e^{5aky}]^{-5/2}$ with $\sigma$ a dimensionless 
constant. Note that when $a=1$ and $\sigma=0$ we recover the usual RS result. However, the choice of $\sigma$ can 
easily modify the warp factor from its conventional behavior. What is the nature of the bulk matter distribution 
that yields this metric? The solution to this can be obtained by considering Eqs.(24) and (26) with $F_{P,Q}=0$. 
Eq.(26) immediately tells us that the trace of the 5-d stress-energy tensor $T=g^{AB}T_{AB}=T_0$, a constant, so that 
if we define $T_{55}=T(y)$ then we must have $T_{\mu\nu}={1\over {4}}e^{-2A}\eta_{\mu\nu}(T_0+T(y))$, with $T(y)$ 
an arbitrary function. The $\mu\nu$ and $55$ components of the equations of motion when combined then provides a  
first order differential equation for $T(y)$ that can be solved in a straightforward manner. Setting $a=5,~b=
(R_0-F_R-5F_0/2)/M^2,~c=(F_0/F_R-R_0)/12,~d=-M^2/6F_R$ and defining $\Delta=bc-ad$ with $x(y)=T(y)/M^5$, we obtain 
the general solution (assuming $A'' \neq 0$)
\begin{equation}
x(y)={-c\over {d}}+{\Delta\over {bd}}~{\rm tanh}^2\Big[\sqrt {1\over {2}}{b\Delta}(y+C)\Big]\,, 
\end{equation}
with $C$ an integration constant. Though this is far from a uniform energy 
distribution (away from the TeV brane) it does 
lead to a space of constant curvature but not one which is maximally symmetric. Thus we see that it is possible that 
the requirement that $R=R_0$ does allow for the possibility of more complex solutions than that employed in the 
original RS model.

\section{Discussion and Conclusions}

In this paper we have begun an examination of how generic higher curvature terms in the gravitational 
action can alter the predictions of both the ADD model and the RS model defined on a interval to avoid possible 
brane singularities. We have assumed that the traditional assumptions of the two models, \eg, SM localized 
matter in a conformally flat bulk, remain valid; we have not considered in detail more complex setups that may now be 
allowed by the modified equations of motion. To be more concrete, we have further assumed that the EH action 
is now generalized to an action which is of the form 
$F(R,P,Q)$ where $F$ is a well-behaved function, $P=R_{AB}R^{AB}$ and $Q=R_{ABCD}R^{ABCD}$. In D-dimensions 
this action results in a propagating massless tensor field (identified with the usual graviton), a massive ghost 
tensor field, as well as a massive (possibly tachyonic) 
scalar. The potentially dangerous ghost is removable from the perturbative 
spectrum, \ie, it becomes infinitely massive, 
if we demand that $F=F(R,Q-4P)$ only. The remaining new scalar field has a bulk mass whose value is naturally 
expected to be of order the fundamental scale, $M$, in either scenario. The resulting ADD and RS models are 
altered in similar ways from their traditional standard forms: 

($i$) New scalar KK excitations appear in the spectrum of both models in a rather benign fashion coupling to 
the trace of the stress-energy tensor of the localized SM fields. Since this trace is proportional to SM 
masses, the couplings of these scalars are relatively strongly suppressed in comparison to those for the KK 
gravitons at typical collider energies in both models. In the ADD model, the KK scalar excitations begin at 
a mass $\sim M \sim$ TeV. Consequently their contributions to missing-energy signatures as well as to the 
usual dimension-8 contact interactions are further kinematically suppressed. Thus at leading order these 
new scalars do not much influence ADD collider signatures. In RS, 
the bulk scalar mass tends to be large so that the lightest scalar KK state is several times more massive than is 
the lightest KK graviton. Given their rather weak couplings such states will be difficult to observe at 
colliders. 

($ii$) The basic model relationships involving the fundamental and 4-d Planck masses in both models get rescaled 
by functions of $F$ and its derivatives evaluated in the corresponding background metric of the two models: 
in ADD we obtain $\mpl^2=V_nM^{n+2}F_R$ while in RS we obtain $\mpl^2=H(k)~M^3/k$ where $H(k)$ is explicitly given in 
Eq.(39). Assuming that $M$ is a fixed fundamental parameter these modifications lead to changes in the graviton KK 
sectors of both models. In the ADD case, since $\mpl$ is known and $M$ is an input parameter for any given $F_R$ 
the volume of the compactified space and, hence, the value of the compactification radius which sets the 
graviton KK mass scale is altered. Due to the presence of the $F_R$ factor the emission rate for gravitons in the 
collisions of SM particles and for the graviton exchange amplitude are both modified by potentially O(1) effects. 
Similarly in RS, $k$ is a 
derived parameter which sets the scale for all the KK states. The constraint Eq.(28) allows us to calculate $k$  
in terms of the input parameter $M$ and the function $F$ thus providing for us with $H(k)$. In a manner 
similar to ADD, the presence of $H$ rescales the coupling strengths of the of the KK graviton states to the SM 
fields thus modifying the widths and production cross sections at colliders by potentially O(1) factors.  

As we have seen, the extension of the EH action to a more complicated structure can lead to significant 
quantitative modifications 
to both the ADD and RS model predictions in the simplest possible case. The observation of such effects at future 
colliders could tell us valuable information about the underlying theory of gravity.

{\it Note Added}: After this paper was essentially completed, Ref.{\cite {Aslan:2006qi}} appeared which discusses  
generalized actions for the ADD model and thus has some common areas with the present work. Where the 
two papers overlap there is general qualitative agreement though the points of view are somewhat different.

\noindent{\Large\bf Acknowledgments}

The author would like to thank J.Hewett for discussions related to this work.

%
\def\MPL #1 #2 #3 {Mod. Phys. Lett. {\bf#1},\ #2 (#3)}
\def\NPB #1 #2 #3 {Nucl. Phys. {\bf#1},\ #2 (#3)}
\def\PLB #1 #2 #3 {Phys. Lett. {\bf#1},\ #2 (#3)}
\def\PR #1 #2 #3 {Phys. Rep. {\bf#1},\ #2 (#3)}
\def\PRD #1 #2 #3 {Phys. Rev. {\bf#1},\ #2 (#3)}
\def\PRL #1 #2 #3 {Phys. Rev. Lett. {\bf#1},\ #2 (#3)}
\def\RMP #1 #2 #3 {Rev. Mod. Phys. {\bf#1},\ #2 (#3)}
\def\NIM #1 #2 #3 {Nuc. Inst. Meth. {\bf#1},\ #2 (#3)}
\def\ZPC #1 #2 #3 {Z. Phys. {\bf#1},\ #2 (#3)}
\def\EJPC #1 #2 #3 {E. Phys. J. {\bf#1},\ #2 (#3)}
\def\IJMP #1 #2 #3 {Int. J. Mod. Phys. {\bf#1},\ #2 (#3)}
\def\JHEP #1 #2 #3 {J. High En. Phys. {\bf#1},\ #2 (#3)}

\end{document}